# Towards Nonadditive Quantum Information Theory

Sumiyoshi Abe[1] and A. K. Rajagopal[2]

[1]*College of Science and Technology, Nihon University,*
*Funabashi, Chiba 274-8501, Japan*

[2]*Naval Research Laboratory, Washington, DC 20375-5320*

**Abstract**.  A definition of the nonadditive (nonextensive) conditional entropy indexed by $q$ is presented. Based on the composition law in terms of it, the Shannon-Khinchin axioms are generalized and the uniqueness theorem is established for the Tsallis entropy. The nonadditive conditional entropy, when considered in the quantum context, is always positive for separable states but takes negative values for entangled states, indicating its utility for characterizing entanglement. A criterion deduced from it for separability of the density matrix is examined in detail by using a bipartite spin-1/2 system. It is found that the strongest criterion for separability obtained by Peres using an algebraic method is recovered in the present information-theoretic approach.

In this contribution, we report recent results [1] in developing a nonadditive generalization of quantum information theory based on the idea of Tsallis' entropy [2]. In the field of mathematical information theory, the entropy similar to Tsallis' has been known for a long time [3]. It seems, however, that it has not been fully developed due to the lack of the concept of the $q$-expectation value [4] discussed below. The introduction of the $q$-expectation value enables us to construct the important notion of the nonadditive conditional entropy. As we shall see, such a theory is, when generalized to include quantum theory, very useful in the context of quantum information. It is shown that this generalization leads to characterization of quantum entanglement and yields the separability criterion for the density matrix.

First, let us recall the Boltzmann-Shannon entropy

$$S[p] = S(p_1, p_2, \ldots, p_W) = -\sum_{i=1}^{W} p_i \ln p_i, \qquad (1)$$

where $\{p_i\}_{i=1,2,\ldots,W}$ is a normalized probability distribution and $W$ is the number of microscopic states. This entropy is characterized by the Shannon-Khinchin axioms, which are given as follows [5]:

[I]   $S(p_1, p_2, \ldots, p_W)$ is continuous with respect to all its arguments
      and takes its maximum for the equiprobability distribution $p_i = 1/W$
      $(i = 1, 2, \ldots, W)$,

[II]  $S[A, B] = S[A] + S[B|A]$,

[III]     $S(p_1, p_2, \ldots, p_W, 0) = S(p_1, p_2, \ldots, p_W)$.

In the second axiom, $S[A, B]$ and $S[A]$ are the entropies of the composite system $(A, B)$ with the joint probability distribution $p_{ij}(A, B)$ $(i = 1, 2, \ldots, W; j = 1, 2, \ldots, W')$ and the subsystem $A$ with the marginal probability distribution $p_i(A) = \sum_{j=1}^{W'} p_{ij}(A, B)$, respectively. $S[B|A]$ is the conditional entropy associated with the conditional probability distribution $p_{ij}(B|A) = p_{ij}(A, B)/p_i(A)$:

$$S[B|A] = \langle S[B|A_i] \rangle^{(A)} \equiv \sum_{i=1}^{W} p_i(A) \, S[B|A_i], \qquad (2)$$

where $S[B|A_i]$ is the entropy of the conditional probability distribution. In the special case when $A$ and $B$ are statistically independent, then, $S[B|A] = S[B]$, and therefore the following law of additivity holds:

$$S[A, B] = S[A] + S[B]. \qquad (3)$$

The uniqueness theorem [5] tells us that a quantity satisfying the axioms [I]-[III] is unequivocally the Boltzmann-Shannon entropy in eq. (1), up to a constant multiplicative factor. We point out that there is a correspondence relation between the Bayes multiplication law and the axiom [II]:

$$p_{ij}(A, B) = p_i(A) p_{ij}(B|A) \leftrightarrow S[A, B] = S[A] + S[B|A]. \tag{4}$$

Now, we consider a nonadditive extension of the above structure based on the Tsallis entropy indexed by a positive parameter $q$

$$S_q[p] = S_q(p_1, p_2, \text{L}, p_W) = \frac{1}{1-q} \left[ \sum_{i=1}^{W} (p_i)^q - 1 \right]. \tag{5}$$

This quantity converges to the Boltzmann-Shannon entropy $S[p]$ in the limit $q \to 1$. It is positive and concave, and satisfies the H-theorem, just as $S[p]$. It also fulfills the axioms [I] and [III] for the Boltzmann-Shannon entropy. However, the additivity indicated in eq. (3) is violated, in general. To exhibit this feature, it is commonly stated in the literature of nonextensive statistical mechanics that the factorized joint probability distribution $p_{ij}(A, B) = p_i(A) p_j(B)$ gives rise to pseudoadditivity

$$S_q[A, B] = S_q[A] + S_q[B] + (1-q) S_q[A] S_q[B]. \tag{6}$$

However, this discussion is unsatisfactory, because the assumption of the factorization of the joint probability distribution is not justified in nonextensive systems. For example, if the subsystems are interacting through a long-range force, their spatial separation does not make them independent, since the effect of the interaction persists at all distances. In fact, it turns out that the maximum entropy probability distribution obtained from the Tsallis entropy with constraints on physical quantities never factorizes [6]. On the other hand, the Bayes multiplication law always holds in any situation. This consideration leads us to seek a definition of the nonadditive conditional entropy.

A definition of the nonadditive conditional entropy we propose here is

$$S_q[B|A] = \langle S_q[B|A_i]\rangle_q^{(A)} \equiv \frac{\sum_{i=1}^{W}[p_i(A)]^q S_q[B|A_i]}{\sum_{i=1}^{W}[p_i(A)]^q}. \tag{7}$$

This is the $q$-expectation value [4] of the Tsallis entropy $S_q[B|A_i]$ of the conditional probability distribution and is a natural generalization of eq. (2). An advantage in this definition may be seen as follows. Using eq. (5), we can rewrite eq. (7) as

$$S_q[B|A] = \frac{S_q[A,B] - S_q[A]}{1 + (1-q)S_q[A]}. \tag{8}$$

Rearranging this equation, we find

$$S_q[A,B] = S_q[A] + S_q[B|A] + (1-q)S_q[A]S_q[B|A]. \tag{9}$$

Therefore, we observe the correspondence relation

$$p_{ij}(A,B) = p_i(A)p_{ij}(B|A)$$

$$\leftrightarrow S_q[A,B] = S_q[A] + S_q[B|A] + (1-q)S_q[A]S_q[B|A], \tag{10}$$

which is a natural nonadditive generalization of eq. (4).

In a recent work [7], the following axioms were presented:

[I]*   $S_q(p_1, p_2, \text{L}, p_W)$ is continuous with respect to all its arguments

and takes its maximum for the equiprobability distribution $p_i = 1/W$

$(i = 1, 2, \text{L}, W)$,

[II]*   $S_q[A, B] = S_q[A] + S_q[B|A] + (1-q) S_q[A] S_q[B|A]$,

[III]*  $S_q(p_1, p_2, \text{L}, p_W, 0) = S_q(p_1, p_2, \text{L}, p_W)$.

It was shown [7] that a quantity satisfying these axioms leads uniquely to the Tsallis entropy in eq. (5).

Let us reconsider the above discussion in the quantum mechanical context. In quantum information theory, it is of central importance to understand the properties of quantum entanglement. We demonstrate that the nonadditive conditonal entropy will lead us to characterization of quantum entanglement in the information-theoretic language. The probability distribution is now replaced by the normalized density matrix $\hat{\rho}$. The quantum counterpart of eq. (5) for the composite system $(A, B)$ is

$$S_q[\hat{\rho}(A, B)] = \frac{1}{1-q} \left[ \text{Tr} \left( \hat{\rho}(A, B) \right)^q - 1 \right], \tag{11}$$

where Tr stands for the trace operation over the composite system. This quantity converges to the celebrated von Neumann entropy in the limit $q \to 1$. To construct the nonadditive quantum conditional entropy, we substitute the Tsallis entropy of the marginal density matrix, $\hat{\rho}(A) = \text{Tr}_B \, \hat{\rho}(A, B)$ with $\text{Tr}_B$ standing for the partial trace operation over the subsystem $B$, as well as eq. (11) into eq. (8):

$$S_q[B|A] = \frac{S_q[\hat{\rho}(A, B)] - S_q[\hat{\rho}(A)]}{1 + (1-q) S_q[\hat{\rho}(A)]}. \tag{12}$$

In the limit $q \to 1$, this quantity approaches the conditional von Neumann entropy, which has been discussed in Ref. [8] in the context of separability of the density matrix.

It should be noted that though classically the nonadditive conditional entropy is always nonnegative, its quantum counterpart in eq. (12) can be negative. This can be easily seen if $\hat{\rho}(A, B)$ is taken to be a pure state. In this case, $S_q[\hat{\rho}(A, B)]$ vanishes, whereas $S_q[\hat{\rho}(A)]$ is positive in general. Special is the case when $\hat{\rho}(A)$ is also a pure state and $S_q[\hat{\rho}(A)]$ vanishes. In a more general situation in which the state is mixed, $S_q[B|A]$ vanishes for the product state $\hat{\rho}(A, B) = \hat{\rho}(A) \otimes \hat{\rho}(B)$. These facts show that the change of sign of the nonadditive quantum conditional entropy gives rise to a criterion of separability of quantum states. To be explicit, let us consider the classically correlated state, which is also referred to as the separable state in the literature. This state is defined by a convex combination of product states [9]:

$$\hat{\rho}(A, B) = \sum_\lambda w_\lambda \, \hat{\rho}_\lambda(A) \otimes \hat{\rho}_\lambda(B), \tag{13}$$

where $0 \leq w_\lambda \leq 1$ and $\sum_\lambda w_\lambda = 1$. It is known that this state can be modeled by using local hidden-variable theories and satisfies the Bell inequalities. To calculate the nonadditive quantum conditional entropy of this state, we employ the spectral resolutions of $\hat{\rho}_\lambda(A)$ and $\hat{\rho}_\lambda(B)$:

$$\hat{\rho}_\lambda(A) = \sum_a p_\lambda(a) |a\rangle\langle a|, \qquad \hat{\rho}_\lambda(B) = \sum_b r_\lambda(b) |b\rangle\langle b|, \qquad (14)$$

where $\{|a\rangle\}$ and $\{|b\rangle\}$ are respectively the orthonormal eigenstates of the subsystems $A$ and $B$, $0 \leq p_\lambda(a), r_\lambda(b) \leq 1$, and $\sum_a p_\lambda(a) = \sum_b r_\lambda(b) = 1$. Then, the nonadditive quantum conditional entropy is calculated to be

$$S_q[B|A] = \frac{\sum_a \left[\sum_\lambda w_\lambda p_\lambda(a)\right]^q S_q[B|a]}{\sum_a \left[\sum_\lambda w_\lambda p_\lambda(a)\right]^q}, \qquad (15)$$

where

$$S_q[B|a] \equiv \frac{1}{1-q}\left\{\sum_b [\pi(b|a)]^q - 1\right\}, \qquad (16)$$

$$\pi(b|a) \equiv \frac{\sum_\lambda w_\lambda p_\lambda(a) r_\lambda(b)}{\sum_\lambda w_\lambda p_\lambda(a)}. \qquad (17)$$

Clearly, $0 \leq \pi(b|a) \leq 1$ and $\sum_b \pi(b|a) = 1$, and therefore, $S_q[B|a]$ is nonnegative. Thus, we see that the nonadditive quantum conditional entropy is nonnegative for any classically correlated state.

Now, let us consider a simple bipartite spin-$1/2$ system (i.e., a $2 \times 2$ system) to see if the nonadditive quantum conditional entropy can yield a criterion for separability of the density matrix. Our primary purpose is to clarify how the separability criterion can be strengthened by controlling the value of the parameter $q$. Consider a parametrized form of the Werner-Popescu state [9,10]

$$\hat{\rho}(A, B) = \frac{1-x}{4} \hat{I}_A \otimes \hat{I}_B + x |\Psi^-\rangle\langle\Psi^-|, \tag{18}$$

where $0 \leq x \leq 1$, $\hat{I}_A$ ($\hat{I}_B$) the $2 \times 2$ identity matrix in the space of spin $A$ ($B$), and $|\Psi^-\rangle$ the singlet state given by

$$|\Psi^-\rangle = \frac{1}{\sqrt{2}} \left( |\uparrow\rangle_A \otimes |\downarrow\rangle_B - |\downarrow\rangle_A \otimes |\uparrow\rangle_B \right). \tag{19}$$

After some algebra, the nonadditive quantum conditional entropy is found to be

$$S_q[B|A] = S_q[A|B] = \frac{1}{1-q} \left[ \frac{3}{2}\left(\frac{1-x}{2}\right)^q + \frac{1}{2}\left(\frac{1+3x}{2}\right)^q - 1 \right]. \tag{20}$$

This is a monotonically decreasing function of $x$ for a fixed value of $q$. The minimum value of $x$ at which $S_q[B|A]$ changes its sign is

$$x = \frac{1}{3}, \qquad (21)$$

which is reached in the limit $q \to \infty$. This means that the Werner-Popescu state with $x \in [1/3, 1]$ is not separable, that is, entangled. Local realism does not hold in this range of the parameter. It can be shown that the conditional von Neumann entropy, $S[B|A] := \lim_{q \to 1} S_q[B|A]$, changes its sign at $x \cong 0.748$. Also, it is known that the discussion of the Bell inequalities yields $x = 1/\sqrt{2} \cong 0.707$. Therefore, we see that the nonadditive quantum conditional entropy leads to a criterion which is much stronger than those derived from the Bell inequalities and the conditional von Neumann entropy. Actually, the value in eq. (21) has been derived in Ref. [11] using the algebraic method of partial transposition of the density matrix and is known to be the strongest criterion [12]. Here, we have derived this criterion based entirely on the information-theoretic approach.

In conclusion, we have presented a definition of the nonadditive (nonextensive) conditional entropy indexed by $q$. Based on the composition law in terms of it, we have generalized the Shannon-Khinchin axioms for the Tsallis entropy. Furthermore, we have discussed the nonadditive conditional entropy in the quantum context. We have derived the strongest criterion for separability of the density matrix of a bipartite spin-1/2 (i.e., $2 \times 2$) in the limit $q \to \infty$. It is known [12] that the method of Ref. [11] gives the strongest criteria only for $2 \times 2$ and $2 \times 3$ systems. Therefore, it is of interest to examine the present nonadditive quantum conditional entropy for $3\times 3$ or other general systems. The work in this direction is in progress.

We thank Professors Paolo Grigolini and Constantino Tsallis for kindly inviting us to the workshop. We also thank Dr. R. W. Rendell for his help in numerical determination of the zero of the conditional von Neumann entropy. One of us (S. A.) is supported by the GAKUJUTSU-SHO Program of College of Science and Technology, Nihon University. He acknowledges the warm hospitality of Naval Research Laboratory, Washington, DC, extended to him. The other author (A. K. R.) is supported in part by the US Office of Naval Research.